\documentclass[graybox]{svmult}

\usepackage{helvet}         
\usepackage{courier}        
\usepackage{type1cm}        
\usepackage{makeidx}         
\usepackage{graphicx}        
\usepackage{multicol}        
\usepackage[bottom]{footmisc}

\usepackage[bookmarks=true,colorlinks=true,linkcolor=black,citecolor=blue,urlcolor=blue,bookmarksnumbered]{hyperref}

\usepackage[utf8]{inputenc}

\usepackage{amsmath,amssymb}
\usepackage[colorlinks=true]{hyperref}
\usepackage{dsfont}
\usepackage{xcolor}
\usepackage{bm}

\usepackage{subfigure}
\usepackage{tikz}

\usepackage{bm}
\usepackage{multirow}
\usepackage{enumitem}

\newcommand{\fk}[1]{\mathfrak {#1}}

\newcommand{\rn}[1]{\MakeUppercase{\romannumeral #1}}

\begin{document}

\title*{Automorphic symmetries and $ AdS_{n} $ integrable deformations}
\author{Anton Pribytok}
\institute
{
Anton Pribytok \at Hamilton Mathematics Insitute, Trinity College Dublin, \email{apribytok@maths.tcd.ie}
} 
\maketitle

\abstract{We develop technique based on boost automorphism for finding new lattice integrable models with various dimensions of local Hilbert spaces. We initiate the method by implementing it to two-dimensional models and resolve classification problem, which not only confirms known vertex model solution space, but also extends to the new $ \fk{sl}_{2} $ deformed sector. The generalisation of the approach for string integrable models is provided and allows to find new integrable deformations and associated $ R $-matrices. Hence our new integrable solutions appear to be of non-difference form that admit $ AdS_{2} $ and $ AdS_{3} $ $ S $-matrices as special cases, we also obtain embedding of double deformed sigma model $ R $-matrix into our solution. The braiding and crossing for the novel models as well as their emergence with respect to the deformation parameter $ k $ are shown. 	}

\vspace{4mm}
The present contribution is based on series of works \cite{DeLeeuw:2019gxe,DeLeeuw:2020ahx,deLeeuw:2020xrw,deLeeuw:2021ufg} and relevant to the questions discussed below.
\vspace{-4mm}

\section{The Method: Automorphic Symmetry}
\label{sec:1}

\textbf{Introduction}. \hspace{3mm} An integrable spin chain is characterised by the hierarchy of mutually commuting conserved quantities, \textit{e.g.} charge operators $ \mathbb{Q}_{2} $, $ \mathbb{Q}_{3} $, $ \mathbb{Q}_{4} $, $ \dots $, $ \mathbb{Q}_{r} $ where $ r $ denotes interaction range. In this respect, one considers local vector space $ \mathfrak{H} \simeq \mathbb{C}^{n} $, $ n=2s+1 $. Then spin-$ s $ chain configuration space is given by the $ L $-fold tensor product
\vspace{-2mm}
\begin{equation}\label{key}
	\text{Complete Fock space:} \quad H = \mathfrak{H}_{1} \otimes \dots \otimes \mathfrak{H}_{L} \equiv \bigotimes \limits_{i=1}^{L}\mathfrak{H}_{i} \qquad \mathfrak{H}_{i} \simeq \mathfrak{H}
\end{equation}
\vspace{-2mm}

We will consider homogeneous periodic integrable spin chains (lattice integrability) with the \textbf{N}earest-\textbf{N}eighbour interaction (NN). In magnon propagating systems by definition one can obtain an analogue of momentum operator (shifts) $ \mathbb{Q}_{1} \equiv P $ and 2-site NN charge -- Hamiltonian $ \mathbb{Q}_{2} \equiv H $.
Quantum integrability of such a system is now characterised by \textit{quantum} R-matrix that satisfies \textit{quantum} \textit{\textbf{Y}ang}-\textit{\textbf{B}axter} \textit{\textbf{E}quation} \cite{Yang:1966a,Yang:1967a,Baxter:1972a,Jimbo:1989a}
\begin{equation}
	R_{12}(u,v)R_{13}(u,w)R_{23}(v,w) = R_{23}(v,w)R_{13}(u,w)R_{12}(u,v) 
\end{equation}
where $ R $-matrix properties and structure imply underlying integrability of the model. $ R $-matrix is an operator on spectral parameters associated to the local spaces, specifically for a number of integrable classes such dependence is of difference (or additive) form:
\begin{equation}\label{key}
	R_{ab}(u,v) \rightarrow R_{ab}(u-v)
\end{equation}

$ \mathcal{B} $\textbf{oost operator}.\hspace{2mm} In general, in order to obtain conserved charges, one needs to define $ R $-monodromy and apply $ \log $-derivative
\begin{equation}\label{key}
	\mathbb{Q}= \sum_{n=1}^L R_{n,n+1}^{-1}(0) \frac{d}{du}R_{n,n+1} \equiv \sum_n \mathcal{Q}_{n,n+1}
\end{equation}
with $ \mathcal{Q} $ to denote local densities. Range $ r $ local density can be defined as
\begin{equation}\label{Q_3}
	\mathbb{Q}_{r} \equiv \sum_n \mathcal{Q}_{n,n+1,\dots,n+r-1}
\end{equation}
however a method is required to construct all higher charges in the commuting hierarchy. By Master Symmetries it is possible to show that the tower of conserved charges $ \mathbb{Q}_{r} $, $ r = 2, 3, 4, \dots n $ can be recursively generated by the \textit{boost operator} $ \mathcal{B} \left[ \mathbb{Q}_{2} \right] $ \cite{Tetelman:1982a,Loebbert:2016a}
	\begin{equation}\label{Boost}
		\mathcal{B}[ \mathbb{Q}_{2}] \equiv \sum_{k=-\infty}^\infty k \mathcal{H}_{k,k+1} \quad \rightarrow \quad \mathbb{Q}_{r+1} \simeq [ \mathcal{B}[ \mathbb{Q}_2 ],\mathbb{Q}_r]
	\end{equation}
with Hamiltonian density $ \mathcal{H}_{k,k+1} $. Strictly, the boost operator $ \mathcal{B} \left[ \cdot \right] $ is well-defined on \textit{infinite length spin chains}, which contrasts our closed spin chain system. However the $ \mathcal{B} $-generated conserved charges and their commutators represent finite range operators, so that such construction can be shown to be consistently implemented on spin chains with periodic conditions.

To address the new models, one can demand a generic ansatz for $ \mathcal{H} $ based on an appropriate representation of the underlying symmetry algebra and consistent generator basis \cite{DeLeeuw:2019gxe}. Specifically, starting from $ \mathcal{Q}_{ij} = \mathcal{H}_{ij} \equiv A_{\mu \nu} \hspace{1mm} \sigma^{\mu} \otimes \sigma^{\nu} $ ansatz and commutation of $ \mathbb{Q}_{r} $ charges with finite range tensor embedding $ \sigma^{\mu}_{k} = \mathds{1} \otimes \dots \underbrace{ \otimes \, \sigma^{\mu} \, \otimes}_{\text{k}} \dots \otimes \mathds{1} $, as well as use of symmetry algebra, we can obtain an algebraic system on the $ A_{\mu \nu} $ coefficients. In general, $ [ \mathbb{Q}_{r}, \mathbb{Q}_{s} ] $ commutator provides $ \frac{1}{2}(3^{r+s-1}-1) $ polynomial equations of degree $ r + s - 2 $ \cite{Bargheer:2008a}. To note, that the existence of the first commutator for the studied systems was sufficient to completely resolve for the Hamiltonian. However analytic (recursive) proof of the sufficiency of $ \left[ \mathbb{Q}_{2}, \mathbb{Q}_{3} \right] = 0 $ is still not present for all integrable sectors \cite{Grabowski:1995a}.

Moreover a set of integrable transformations \cite{DeLeeuw:2019gxe} is needed to reduce the complete solution space to the characteristic generators, which allow to show all distinct integrable classes. The necessary and sufficient transformation symmetries include: Norms and shifts of $ R $, Reparametrisation $ R(\mathfrak{f}(u),\mathfrak{f}(v)) $, Local Basis Transform $ R^{\mathcal{V}}(u,v) = \left[ \mathcal{V}(u) \otimes \mathcal{V}(v) \right] R(u,v) \left[ \mathcal{V}(u) \otimes \mathcal{V}(v) \right]^{-1} $, Discrete Transform of $ PRP $-type, Twisting $ \left[ \mathfrak{T}_{1}(u) \otimes \mathfrak{T}_{2} (v)\right] R \left[ \mathfrak{T}_{2}(u) \otimes \mathfrak{T}_{1} (v)\right] ^{-1} $ as a part of the $ R $-matrix symmetry $ \left[ \mathfrak{T}_{1,2} \otimes \mathfrak{T}_{1,2} , R \right] = 0 $.

\vspace{3mm}
\textbf{Bottom-up construction: $ R $-matrix}. \hspace{3mm} The next goal is to construct $ R $-matrix to each associated generating Hamiltonian $ \mathcal{H} $ (class), for that one needs first to consider expansion of the $ R $ matrix
\begin{equation}\label{R-exp}
	R = P + P\mathcal{H} u + \sum_{ n \geq 2 } R^{\left( n \right)} u^{n}
\end{equation}
where $ P $ is permutation operator. If one substitutes $ R $-matrix Ansatz to YBE, one could potentially solve it recursively for the coefficients in the expansion, but in fact, in a number of cases it becomes impossible to identify the right expansion sequence. Instead the \textit{Hamilton-Cayley} theorem argument can be imposed on the $ R $-expansion \ref{R-exp} with the specific set of functional constraints, which led to $ R \equiv R \left( \mathds{1}, \mathcal{H}, \mathcal{H}^{2}, \mathcal{H}^{3} \right)$ resolution. With this construction one can find all $ \mathbb{C}^{2} $ integrable $ R $-matrices, however in higher dimensions $ \mathbb{C}^{n} $ and arbitrary spectral dependence a stronger generalised approach for finding the $ R $-matrix is proven.
%


$ \fk{sl}_{2} $ \textbf{sector.} \hspace{3mm} As a quick test of the technique provided above, one can consider models with two-dimensional local space $ \mathbb{C}^{2} $ and generic ansatz for $ \mathcal{H} $. It turned not only to show full agreement with the set of integrable models that are found from the YBE resolution (\textit{i.e.}  Heisenberg class, *-magnets, multivertex etc), but also find new higher parametric integrable models in the $ \fk{sl}_{2} $ sector. Some of these new classes exhibit \textit{non-diagonalisability} and \textit{nilpotency} of the $ \mathcal{H} $, but others develop conserved charges with \textit{non-trivial Jordan blocks}, which leads to important results and corollaries. Some $ R_{X} $ matrices from characteristic $ X $ classes include
\begin{align}\label{R_1}
	R_{\text{\rn{1}}}(u) =
	\scriptsize{\begin{pmatrix}
			1 & \frac{a_1 (e^{a_5 u}-1)}{a_5}&  \frac{a_2(1-e^{-a_5 u}) }{a_5} &  \frac{a_1 a_3+a_2a_4}{a_5^2}(\cosh (a_5 u) -1) \\
			0 & 0 & e^{-a_5 u} & \frac{a_4(1-e^{-a_5 u}) }{a_5} \\
			0 & e^{a_5 u} & 0 & \frac{a_3 (e^{a_5 u}-1)}{a_5} \\
			0 & 0 & 0 & 1
	\end{pmatrix}}
\end{align}
or another extended model, which in the parametric reduction limit admits the \textit{Kulish-Stolin} model\footnote{Known one-parameter family of $ \fk{sl}_{2} $ models, which also could be related to \cite{Alcaraz:1994a}. In the past it was conjectured, that higher-parametric generalisations might exist, however it appeared non-resolvable by $ RTT $-approach, and in our computation we prove their existence and relations.}
\begin{equation}
	R_{\text{\rn{6}}}(u)= \scriptsize{\text{$ (1-a_1 u)(1+2a_1 u) $}
		\begin{pmatrix}
			1 & a_2 u & a_2 u & -a_2^2 u^2(2 a_1 u+1 ) \\
			0 & \frac{2 a_1 u}{2 a_1 u+1} & \frac{1}{2 a_1 u+1} & -a_2 u \\
			0 & \frac{1}{2 a_1 u+1} & \frac{2 a_1 u}{2 a_1 u+1} & -a_2 u \\
			0 & 0 & 0 & 1
	\end{pmatrix}}
	\normalsize
\end{equation}
It is important to find all underlying Yangian deformations $ \mathcal{Y}^{*}[\fk{sl}_{2}] $ and associated quantum groups. Also known that Belavin-Drinfeld cohomological classification of quantum symmetries for the novel models is an important question on its own.

\vspace{-4mm}
\section{$ \mathcal{B} $ in $ AdS $ integrability}
\vspace{-1mm}
\label{sec:2}

$ AdS/CFT $ integrability \cite{Maldacena:1999a,Witten:1998a,Klebanov:2004a,Beisert:2012a} implies agreement of global symmetries on both sides of the correspondence, e.g. $ \mathcal{N}=4 $ superconformal symmetry and $ AdS_{5} \times S^{5} $ superspace isometries are described by covering supergroup $ \widetilde{PSU}(2,2 \vert 4) $. The corresponding worldsheet model ($ \sigma $-models) integrability is based on $ \fk{psu}(2,2 \vert 4) $ Lie superalgebra and its broken versions. In this setting, the scattering process is described by the $ S $- or $ R $-matrix with arbitrary dependence on the spectral parameter.

$ \mathcal{B} $ \textbf{generalisation.} \hspace{3mm} To address novel results in AdS string background sector we would need to develop boost automorphism  method for operators with generic spectral dependence \cite{deLeeuw:2020ahe}, which intermediately will result in non-additive form of YBE. One will be able to obtain nontrivial constraint system from the commuting tower of the new $ \mathbb{Q}_{r} $ charges
\begin{equation}
	\mathcal{B}\left[ \mathbb{Q}_{2} \right] = \sum_{k=-\infty}^{+\infty} k \mathcal{H}_{k,k+1}(\theta) + \partial_{\theta} \qquad \mathbb{Q}_{r+1} = \left[ \mathcal{B}\left[ \mathbb{Q}_{2} \right], \mathbb{Q}_{r} \right]\quad r>1
\end{equation}
\begin{equation}
	\left[ \mathbb{Q}_{r+1},\mathbb{Q}_{2} \right] \Rightarrow \left[ \left[ \mathcal{B} [\mathbb{Q}_2],\mathbb{Q}_r \right],\mathbb{Q}_{2} \right] + \left[ d_{\theta}\mathbb{Q}_r,\mathbb{Q}_{2} \right] = 0
\end{equation}
from here follows the first order nonlinear ODE coupled system. For the trigonometric and elliptic sectors in $ AdS $ spin chain picture generic ansatz would be
\begin{align}\label{Az8v}
	\begin{split}
		\mathcal{H}  =\,& 
		h_1  \text{ }\mathds{1} + h (\sigma _z\otimes \mathds{1}- \mathds{1}\otimes \sigma _z) + h_3  \sigma _+\otimes \sigma _-  + h_4 \sigma_-\otimes \sigma _+ + \\
		&     h_5
		( \sigma _z \otimes \mathds{1} +  \mathds{1} \otimes \sigma _z ) + h_6 \sigma _z\otimes \sigma _z  +
		h_7 \sigma _-\otimes
		\sigma _- + h_8 \sigma _+\otimes \sigma _+
	\end{split}
\end{align}
with $ h_{k}  $ to be a function on the spectral parameter and $ \sigma^\pm=\dfrac{1}{2}(\sigma_x\pm i \sigma_y) $.

\vspace{3mm}
\textbf{Sutherland type couple.} \hspace{3mm} To obtain R-matrix in such setting, one can expand YBE to the first order, associate spectral parameters along with $ R $-matrix identifications, which will result in coupled differential $ R $-system
\begin{equation}
	\begin{cases}
		\left[ R_{13}R_{23}, \mathcal{H}_{12}(u)\right] = (\partial_{u}R_{13})R_{23} - R_{13} (\partial_{u} R_{23})  \qquad u_{1}=u_{2} \equiv u\\
		\left[ R_{13}R_{12}, \mathcal{H}_{23}(v)\right] = (\partial_{v}R_{13})R_{12} - R_{13} (\partial_{v} R_{12}) \qquad u_{2}=u_{3} \equiv v
	\end{cases}
\end{equation}
which appears sufficient to fix $ R_{ij} = R_{ij}(u,v) $ and equations of the system constitute a reduction from the \textit{Sutherland} equation.

\vspace{3mm}
\textbf{6v: Trigonometric.} \hspace{3mm} We find two classes out of four, whose $ R $-matrices exhibit completely arbitrary spectral dependence and provide deformations relevant for $ AdS_{3} $ models (incl. specific worldsheet model deformations).

\textbf{8v: Elliptic.} \hspace{3mm} Also in elliptic sector we find two novel 8-vertex classes

\vspace{2mm}
8-vertex A class 
\scriptsize
\begin{equation}
	\begin{aligned}
		& r_1(u,v) = r_4(u,v) = \text{sn}(\eta + z) \\[2ex] 
		& r_5(u,v) = r_6(u,v) = \text{sn}(\eta)
	\end{aligned}
	\qquad
	\begin{aligned}
		& r_3(u,v) = r_2(u, v) = \text{sn}(z) \\[2ex]
		& r_7(u,v) = r_8(u,v) = k \text{sn}(\eta) \text{sn}(z) \text{sn}(\eta + z)
	\end{aligned}
\end{equation}
\normalsize

8-vertex B class 
\scriptsize
\begin{equation}
	\begin{aligned}
		& r_1(u,v) = \frac{1}{\Sigma}\left(\sin\eta_+\frac{\text{cn}}{\text{dn}}-\cos\eta_+ \text{sn}\right) \\[2ex]
		& r_3(u,v) = \frac{\pm 1}{\Sigma}\left(\cos\eta_- \text{sn}-\sin\eta_-\frac{\text{cn}}{\text{dn}}\right) \\[2ex] 
		& r_5(u,v) = r_6(u,v)=1 
	\end{aligned}
	\qquad\quad
	\begin{aligned}
		& r_2(u,v) = \frac{\pm 1}{\Sigma}\left(\cos\eta_-\text{sn}+\sin\eta_- \frac{\text{cn}}{\text{dn}}\right) \\[2ex] 
		& r_4(u,v) = \frac{1}{\Sigma}\left(\sin\eta_+\frac{\text{cn}}{\text{dn}}+\cos\eta_+ \text{sn}\right) \\[2ex] 
		& r_7(u,v) = r_8(u,v)=k \frac{\text{sn}\,\text{cn}}{\text{dn}}
	\end{aligned}
\end{equation}
\normalsize
where $ r_{i} $ agree with 8v positions of (\ref{Az8v}), the deformation parameter $ k $, arbitrary $ \eta(u) $ in $ \eta_{\pm} \equiv \frac{\eta(u)-\eta(v)}{2} $, $ \Sigma = \sqrt{\sin\eta(u)\sin\eta(v)} $ and Jacobi elliptic functions $ \text{xn} = \text{xn}(u-v,k^{2}) $ to be $ \{ u,v \} $ dependent. Current class appears in the $ AdS_{2} $ integrable background.

\textbf{Free Fermions}. \hspace{3mm} Important to note that these classes also satisfy algebraic integrable constraint -- \textit{Free fermion} condition \cite{DeLeeuw:2020ahx}. The corresponding characteristic constraint $ \left[ r_{1}r_{4} + r_{2}r_{3} -(r_{5}r_{6}+r_{7}r_{8}) \right]^{2} \cdot \left[ r_{1}r_{2}r_{3}r_{4} \right]^{-1} = \mathcal{K} $ implies Baxter condition for $ \mathcal{K} \neq 0 $ or Free fermion when $ \mathcal{K} = 0 $.

\vspace{-4mm}
\section{$ AdS_{2} $ and $ AdS_{3} $ integrable backgrounds}

\vspace{-2mm}
Completeness of $ AdS/CFT $ correspondence requires integrable string backgrounds of distinct dimensionality, which leads to different amounts of preserved supersymmetry and distinct properties of integrable model.

\bm{$ AdS_{3} \times S^{3} \times \mathcal{M}_{4} $}. \hspace{3mm} In our case $ AdS_{3}/CFT_{2} $ \cite{Sfondrini:2014a} defines $ AdS_{3} \times S^{3} \times \mathcal{M}^{4} $ background under two geometries that preserve 16 supercharges
\vspace{-2mm}
\small
\begin{equation*}
	\begin{cases}
	\mathcal{M}^{4} = T^{4}, \text{with } \fk{psu}(1,1\vert2)^{2} \\
	\mathcal{M}^{4} = S^{3} \times S^{1} , \text{with } \fk{d}(2,1;\alpha)^{2} \sim \fk{d}(2,1;\alpha)_{L} \oplus \fk{d}(2,1;\alpha)_{R} \oplus \fk{u}(1)\\
	\end{cases}
	\vspace*{-2mm}
\end{equation*}
\normalsize
$ \alpha $ is related to the relative radii of the spheres. As it was stated, the underlying $ R $-matrix is of trigonometric type and we find novel 6-vertex B type model to constitute same chirality $ AdS_{3} $ Hamiltonian. It can admit either continuous family of deformations (\textit{spectral functional shifts}) when mapped to \textit{6-vB} or \textit{single-parameter elliptic deformation} when mapped to \textit{8-vB}. In the present setting, we also confirm that $ AdS_{3} \times S^{3} \times T^{4} $ $ R\text{-/} S $-matrix \cite{Borsato:2013a,Borsato:2014a} can be obtained from $ AdS_{3} \times S^{3} \times S^{3} \times S^{1} $ \cite{Borsato:2015a} by appropriate limits. Importantly, we also show that our \textit{6-vB} model allows to embed the two-parameter $ q $-deformed $ R $-matrix \cite{Hoare:2015a} that underlies double deformed Metsaev-Tseytlin model \cite{Metsaev:1998a}.

\vspace{1mm}
$ \bm{AdS_{2} \times S^{2} \times T^{6}} $. \hspace{3mm} The $ AdS_{2} \times S^{2} \times T^{6}$ \cite{Hoare:2016a} model contains \scriptsize $ \dfrac{PSU(1,1\vert2)}{SO(1,1) \times SO(2)} $ \normalsize supercoset, with $ \mathbb{Z}_{4} \in \fk{psu}(1,1\vert2) $, which implies classical integrability, but the construction lacks gauge fixing of $ \kappa $-symmetry. Scattering process on this background can be captured by elliptic deformation of the $ R $-matrix. In this case, the new \textit{8-vB} type, which admits \textit{single-parameter} deformation, represents deformation of the (massive) $ AdS_{2} \times S^{2} \times T^{6} $.

\textbf{Crossing symmetry}. \hspace{3mm} Generically individual blocks $ 4 \times 4 $ obey YBE, Crossing symmetry and Braiding unitarity, however it is necessary to show that the full scattering operator respects them as well\footnote{$ 16 \times 16 $ $ S $-/$ R $-matrix in 2-particle representation that is embedded in $ AdS_{\{ 2,3 \}} $ backgrounds}. In the $ k \rightarrow 0 $ the $ AdS_{3} $ \textit{6-vB} satisfies the crossing symmetry, whereas crossing symmetry works for $ AdS_{2} $ in the $ k \rightarrow \infty $ limit and does not hold for the generic $ k $.

\vspace{-3mm}
\section{Conclusions and remarks}
\vspace{-2mm}
We have developed a method that allows to find new integrable models and their deformations without direct resolution of the YBE. To achieve that, one is required to use automorphic symmetry to build conserved charges and apply property of the integrable commuting hierarchy. Application of invariant transformations to the solution space provides solution generators and the corresponding $ R $-matrix is found by bottom-up approach from $ \mathcal{H} $. That leads to new integrable models or specific extensions and demonstrates universality of the method for periodic or infinite open systems described by a variety of symmetry algebras \cite{DeLeeuw:2019fdv, deLeeuw:2021ufg}.

For the $ AdS $ integrability we considered a generic 8-vertex Ansatz of non-difference form and developed boost automorphism technique also for the string integrable sector, which resulted in differential nonlinear ODE problem. Generalised novel models were found, which also admitted embedding of known integrable models in $ AdS_{\{ 2,3 \}} $ space \cite{deLeeuw:2020ahe}, and new constructions in $ AdS_{5} $ . Such models obey all string integrable symmetry constraints and allow for further restrictions on their structure, as well as provide a proposal for the study of higher parametric $ \sigma $-models, their scattering matrices and quantum limits \cite{deLeeuw:2021ufg}.

\begin{acknowledgement}
I would like to thank Marius de Leeuw, Ben Hoare, Alessandro Torrielli, Ana L. Retore, Chiara Paletta and Paul Ryan for their contribution and useful discussions. AP is supported by the Royal Society RGF$\backslash$EA$\backslash$180167.
\end{acknowledgement}

\end{document}